# DAΦNE LIFETIME OPTIMIZATION WITH COMPENSATING WIRES AND OCTUPOLES

C. Milardi, D. Alesini, M.A. Preger, P. Raimondi, M. Zobov, LNF-INFN, Frascati, Italy
D. Shatilov, BINP, Novosibirsk, Russia.

*Abstract*

Long-range beam-beam interactions (parasitic crossings) were one of the main luminosity performance limitations for the lepton Φ-factory DAΦNE in its original configuration. In particular, the parasitic crossings led to a substantial lifetime reduction of both beams in collision. This puts a limit on the maximum storable current and, as a consequence, on the achievable peak and integrated luminosity. In order to mitigate the problem, numerical and experimental studies of the parasitic crossings compensation by current-carrying wires have been done. During the operation for the KLOE experiment two such wires have been installed at both ends of the interaction region. They produced a relevant improvement in the lifetime of the *weak* beam (positrons) at the maximum current of the *strong* one (electrons) without luminosity loss, in agreement with the numerical predictions. The same compensating mechanism has been adopted during the run for the FINUDA experiment as well, with less evident benefits than in the previous case.

The interplay between nonlinearities originating from the beam-beam interaction and the ring lattice has been studied by theoretical simulation and experimental measurements. Compensation procedures have been set up relying on the electromagnetic octupoles installed on both rings and used in addition to wire compensation.

In this paper the parasitic crossings effects in the DAΦNE interaction regions and their compensation by wires and octupoles are described. A detailed theoretical analysis of the interplay about different non-linearities is presented; eventually experimental measurements and observations are discussed.

## INTRODUCTION

The Frascati Φ-factory DAΦNE is an $e^+e^-$ collider operating at the energy of Φ-resonance (1.02 GeV c.m.) [1]. Its best peak luminosity reached so far is $1.6 \times 10^{32}$ cm$^{-2}$s$^{-1}$ with a maximum daily integrated luminosity of about 10 pb$^{-1}$ [2]. Recently the accelerator complex has undergone a major upgrade with drastic change in its interaction regions (IRs) layout [3].

In order to obtain such a high luminosity at low energy high current bunched beams were stored in two colliding rings sharing two IRs, whose only one at a time hosted an experimental detector (see the original layout in Fig. 1).

Usually, the number of adjacent filled buckets is in the range 109÷111 out of 120 available. A short gap is needed for ion clearing in the electron ring. It's worth reminding that in DAΦNE the bunch separation of 2.7 ns is the shortest among all existing colliders and particle factories.

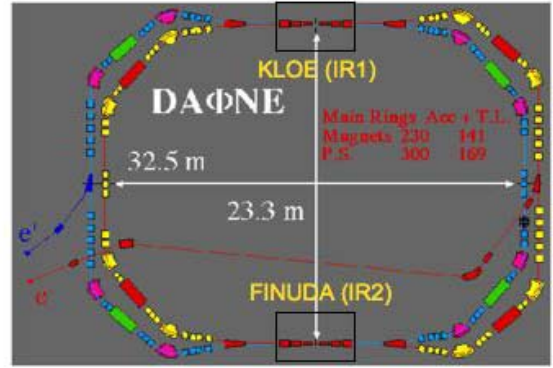

Figure 1: DAΦNE layout before the upgrade.

In order to minimize the effect of parasitic crossings (PC) between the colliding bunch trains the beams collided under a crossing angle in the range 20÷30 mrad. However, despite the crossing angle, the long-range beam-beam interactions (LRBB) remained one of the most severe limitations to the DAΦNE performance in terms of luminosity. In fact LRBB interaction led to a substantial lifetime reduction of both beams, limiting the maximum storable currents and, as a consequence, the maximum achievable peak and integrated luminosity. The latter was strongly influenced by the beam lifetime because in the topping up regime a fraction of the integrated luminosity is lost during the time required to switch the injection system from one beam to the other.

Looking for a compensation scheme to reduce the impact of LRBB interactions, it has been decided to install two windings (wires) at both ends of the IRs. This approach revised an idea originally proposed by J. P. Koutchouk [4] for LHC, and recently tested during single beam operation on SPS [5, 6]. Simulations using LHC compensation devices also predicted a relevant dynamic aperture enlargement for the Tevatron collider [7].

An improvement in the beam lifetime has been also obtained by understanding the interplay between nonlinearities coming from beam-beam interaction and magnetic lattice; the effect has been cured by using the octupole magnets installed on both rings.

The DAΦNE experimental studies about LRBB compensation, using built for the purpose wires and octupoles, yielded quite encouraging results and gave the opportunity, for the very first time, to test the wire compensation scheme in collision.

## PARASITIC CROSSINGS IN IR1

In its original configuration DAΦNE consisted of two independent rings sharing two interaction regions: IR1 and IR2. Bunches experienced 24 PCs in each IR, 12 before and 12 after the main interaction point (IP), until splitter magnets drove them into two different rings.

The KLOE detector was installed in IR1. While delivering luminosity to KLOE [8] bunches collided with a horizontal crossing angle of 29.0 mrad, and were vertically separated in IR2 by a distance larger than 200 $\sigma_y$. For this reason, in the following considerations only LRBB interactions in IR1 are taken into account.

Table 1: Parameters for the Pcs, one every four, in IR1.

| PC order | $Z-Z_{IP}$ [m] | $\beta_x$ [m] | $\beta_y$ [m] | $\mu_x-\mu_{IP}$ | X [$\sigma_x$] | Y [$\sigma_y$] |
|---|---|---|---|---|---|---|
| BB12L | -4.884 | 8.599 | 1.210 | 0.167230 | 26.9050 | 26.238 |
| BB8L | -3.256 | 10.177 | 6.710 | 0.140340 | 22.8540 | 159.05 |
| BB4L | -1.628 | 9.819 | 19.416 | 0.115570 | 19.9720 | 63.176 |
| BB1L | -0.407 | 1.639 | 9.426 | 0.038993 | 7.5209 | 3.5649 |
| IP1 | 0.000 | 1.709 | 0.018 | 0.000000 | 0.0000 | 0.0000 |
| BB1S | 0.407 | 1.966 | 9.381 | 0.035538 | -6.8666 | 3.5734 |
| BB4S | 1.628 | 14.447 | 19.404 | 0.092140 | -16.4650 | 63.196 |
| BB8S | 3.256 | 15.194 | 6.823 | 0.108810 | -18.7050 | 157.74 |
| BB12S | 4.884 | 12.647 | 1.281 | 0.126920 | -22.1880 | 25.505 |

Table 1 summarizes the main parameters for some PCs in the IR1: relative position, beta functions, phase advances with respect to the IP1 and transverse separation in terms of $\sigma_{x,y}$.

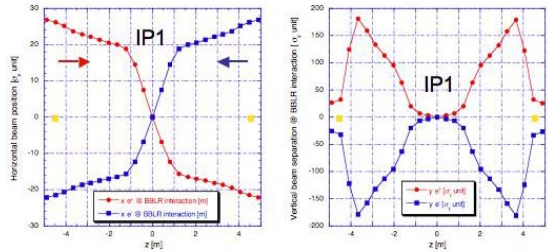

Figure 2: PCs horizontal (left) and vertical (right) beam-beam separation in the IR1 expressed in terms of $\sigma_{x,y}$ and computed for the KLOE optics. Arrows indicate the incoming direction of the positron (red) and electron (blue) beams, yellow dots show the place where the wires are installed.

The more evident effect of the LRBB interactions on the beam dynamics was provided by orbit deflection. In fact, the PCs induce orbit distortion that can be satisfactory reproduced by the machine model, based on the MAD [9] code, when the PCs are taken into account. MAD predictions agree with the orbit distortion obtained from the beam-beam simulation code Lifetrack [10] as well (see Fig. 3).

Moreover, the lifetime of each beam started decreasing during injection of the opposite beam and remained low soon after injection.

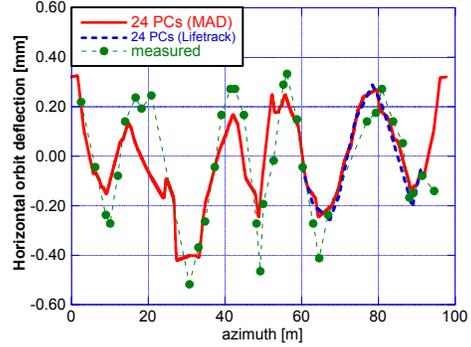

Figure 3: computed orbit deflection due to 24 BBLR interactions for a positron bunch colliding with an electron beam of 10 mA/bunch.

Typically, in collision, the electron beam current reached 1.8÷2.2 A, while the maximum positron beams is 1.3÷1.4 A. Exceeding these values the lifetime of the beams dropped down to 700÷800 sec.

This behaviour has been recognized as one of the main limitations of the collider performance.

## NUMERICAL SIMULATIONS

The "weak-strong" tracking code LIFETRAC was used to simulate the equilibrium distribution of the positron ('weak') beam. The main sources of long beam tails were the 2 PCs nearest to the Main IP, but the other PCs also gave some contribution, so we accounted for all them. The wires were simulated as additional PCs with variable current ("wire-PC"), so that no special tracking algorithm for wires was used.

This approach was justified by the rather large values of the $\beta_{x,y}$ functions at the wire locations (16.5 and 4 m respectively), much larger than both the bunch and wire length. This allowed simulating the interaction with the wire as a single kick, neglecting the effect of displacements of the "strong" bunch: due to synchrotron oscillations the longitudinal coordinate of collision points for the real PC depends on the particle's longitudinal coordinate, while the wire are fixed, but due to the large beta values a shift of few millimetres gives actually no effect. On the other hand, the betas are small enough to produce a large separation in units of the transverse beam size (≈20), so that the actual "shape of wire" (i.e. density distribution inside the wire-PC) can be neglected: it works like a simple 1/r lens. Some simulation results are shown in Fig. 3.

The beam current was chosen to be large enough to yield long beam tails due to PCs (a), then the wires were switched on and the tails reduced (b). When the wires are powered with wrong polarity, the tails blow-up becomes even stronger (c).

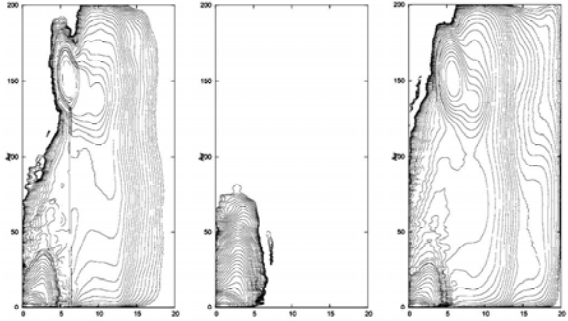

Figure 4: Particle equilibrium density in the normalized transverse phase space, starting from left: wires off (a), wires on (b) and wires powered with wrong polarity (c).

As a matter of fact, the PCs compensation with a single wire on each side of the interaction region was not perfect since distances between the beams at PC locations were different in terms of the horizontal sigma and phase advances between PCs and wires were not completely compensated (see Table 1). Indeed, the numerical simulations did not show improvements in luminosity. However, the positive effect of tails reduction and corresponding lifetime increase is very important, because it leads to a larger integrated luminosity.

## WIRE DESIGN AND INSTALLATION

The wires have been built and installed in IR1 in November 2005. Each device was made by two windings of rectangular shape, 20 coils each, and installed symmetrically with respect to the horizontal plane, see Fig. 3.

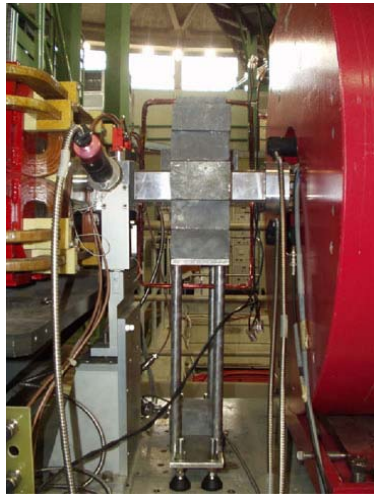

Figure 5: The wires installed at one end of IR1.

Our device differs from the LHC one for several aspects: they were installed outside the vacuum chamber exploiting a short section in IR1, just before the splitters, where the vacuum pipes are separated to host Lambertson type correctors not essential for operation and therefore removed. The wires carried a tunable DC current, and produced a stationary magnetic field with a shape similar to the one created by the opposite beam.

## EXPERIMENTAL RESULTS DURING THE KLOE RUN

A systematic study of the wires in collision has been undertaken during the machine shifts in March 2006.

The wires were powered at 3.6 A to compensate as much as possible the beam lifetime of the positron beam that, due to the limited maximum achievable current [11], can be considered as the 'weak' one.

It has been experimentally verified that the residual orbit distortions with maximum deviations of +0.4 -0.5 mm due to the PCs were very well corrected with wire currents of ≈1 A. This is a proof that the wires behaved as correctors "in phase" with the PCs. It has been also measured that the wires introduced some betatron tunes shifts.

The residual orbit distortion due to the wires at 3.6 A was corrected by the ordinary dipole correctors, while the tune shifts were compensated by means of the quadrupoles in a dispersion free straight section.

Several luminosity runs have been compared switching the wires on and off in order to study their impact on the collisions. In the following the two most relevant sets are presented taking into account 2 hours long runs. The results in Fig. 7 show some clear evidences. Switching on and off the wires we obtain the same luminosity while colliding the same beam currents. The positron lifetime is on average higher when wires are on, while the electron one is almost unaffected. The beam blow-up occurring from time to time at the end of beam injection, corresponding to a sharp increase in the beam lifetime, almost disappears.

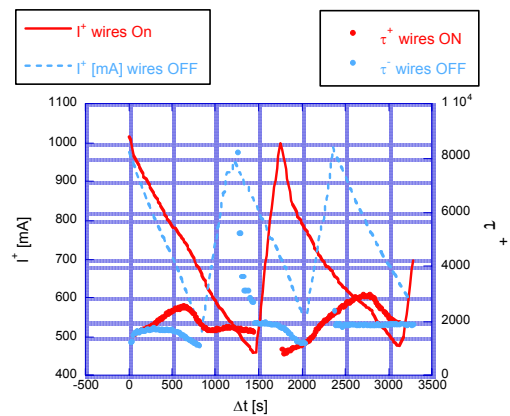

Figure 6: Positron current and lifetime as a function of time: wires on (red) and wires off (cyan).

A further aspect becomes evident when comparing, on the same plot, the positron current and lifetime with and without wires (see Fig. 6). The positron current starts from the same value; then, in the case of wires off, the lifetime of the current is longer than in the case with wires on. In this way it is possible to keep the same integrated luminosity injecting the beam two times only instead of three in the same time interval, or to increase the integrated luminosity by the same factor keeping the same injection rate.

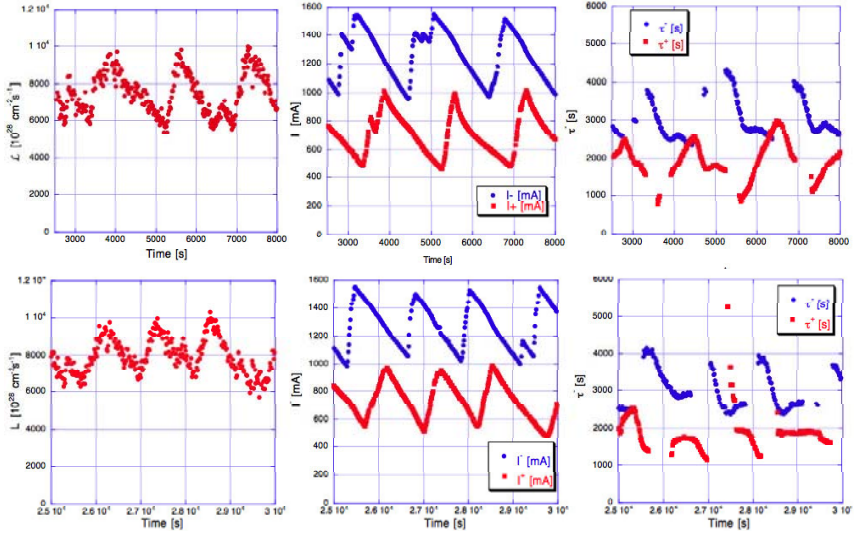

Figure 7: Luminosity, colliding currents and lifetime as a function of time: wires on (upper frames) and wires off (lower frames).

## EXPERIMENTAL RESULTS DURING THE FINUDA RUN

During the upgrade [12] preceding the FINUDA run the KLOE detector has been removed and IR1 replaced with a straight section equipped with four electromagnetic quadrupoles. This was a much more flexible configuration in order to detune the optical functions in the unused IP, and to avoid further contributions to the betatron coupling.

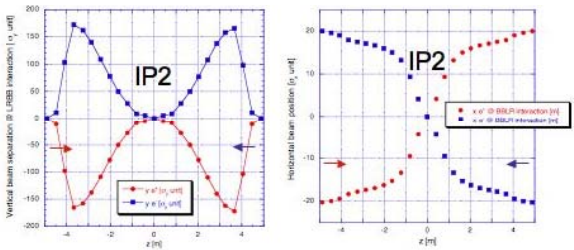

Figure 8: PCs vertical (left) and horizontal (right) beam-beam separation in IR2 expressed in terms of $\sigma_{x,y}$ and computed for the FINUDA optics. Arrows indicate the incoming direction of the positron (red) and electron (blue) beam.

The FINUDA IR was based on four permanent magnet quadrupoles placed inside the FINUDA 1.1 T solenoidal field and on four conventional quadrupoles installed outside it.

Despite the value of the betatron functions at IP2 were the same as during the KLOE run as well as the horizontal crossing angle, due to the different magnetic layout the PCs occurred at smaller beam separation (see Fig. 8) and were more harmful.

Using the wires for LRBB compensation at IR2 produced a few percent increase in the positron lifetime; however, the orbit deflection could not be corrected with a constant current in the wires.

A better compensation of the PCs occurring in IR1 was obtained by halving $\beta_y$ and increasing the beam vertical separation (~2 cm) at IP1, switching on at the same time the wires installed in IR1.

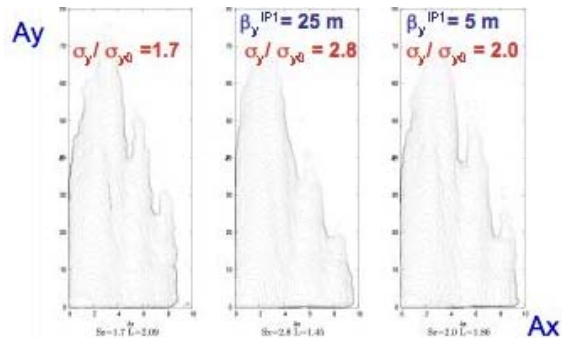

Figure 9: Particle equilibrium density in the normalized transverse phase space computed taking into account the main beam-beam interaction at IP2 (left), adding the contribution of the first PC at IP1, with 2cm vertical separation and a $\beta^{IP1}_y = 25$ m (center), and $\beta^{IP1}_y = 5$ m (right).

Beam-beam simulations, showing the transverse beam blow-up dependence on parasitic crossings, for a given beam-beam separation, have been useful in this optimization process as can be seen from Fig. 9.

## NON-LINEARITY INTERPLAY

LRBB interaction originates nonlinear fields, which interfere and add up with the non-linear terms coming from the ring lattice.

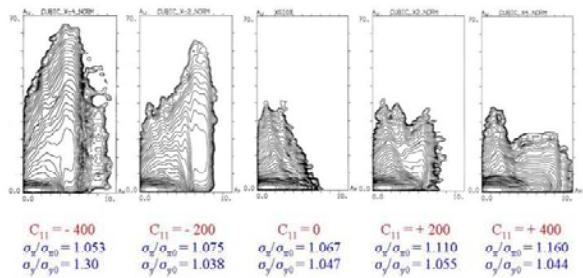

Figure 10: Particle equilibrium density in the normalized transverse phase space computed taking into account the main interaction point and different $C_{11}$ (unit $m^{-1}$) values, for each case the relative variation of the transverse horizontal and vertical dimension are reported.

Such interplay has been studied combining experimental measurements and theoretical simulations [13]. The tune shift on amplitude ($C_{11}$) measurement provided an efficient and simple way to evaluate the lattice contribution to the nonlinear terms for each ring, independently from the influence of PCs.

Simulations have been used to understand the mutual interaction between the two contributions. Fig. 11 shows the growth of the transverse beam dimensions in the case when the main interaction point and the tune shift on amplitude are taken into account. An evident transverse beam blow-up appears when $|C_{11}| > 200$ $m^{-1}$.

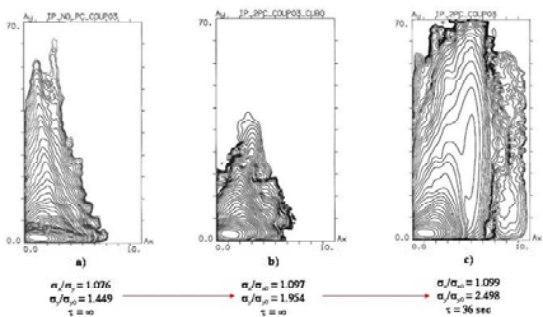

Figure 11: Particle equilibrium density in the normalized transverse phase space computed taking into account the main beam-beam interaction at IP2 (left), adding the contribution of the first PC at IP1, with 2cm vertical separation and a $\beta^{IP1}_y = 25$ m (center), and $\beta^{IP1}_y = 5$ m (right).

When also the first PC was considered the growth in the beam transverse dimensions became even more evident, see Fig. 11, especially for $C_{11} > 0$ affecting mainly the transverse vertical plane.

Unlike the strength of the nonlinear component coming from the lattice, which is fixed, the one due to the PCs depends on the current stored in the colliding bunches. As a result the overall nonlinear term affecting the beam dynamics can have a considerably excursion as the colliding currents decay. Such effect has been clearly observed especially during the last FINUDA run mostly for the positron beam, which at the maximum current, just after injection, had a very low lifetime, less than 500 s. By tuning the working point and the electromagnetic octupole, during the injection, it has been possible to double the positron beam lifetime at its maximum current. The working point was moved toward the integer and the octupole current increased consistently with nonlinearities compensation.

## PARASITIC CROSSINGS IN THE UPGRADED DAΦNE RINGS

Relying on the experience gained about LRBB compensation during the KLOE and FINUDA runs the two DAΦNE IRs have been modified in view of the SIDDHARTA experiment run [14, 15], which will be also used to test a new collision scheme based on large Piwinski angle and *crab-waist* [17]. The vacuum pipe [16] in the unused IR2 provides now complete beam separation while the one in IR1 consists of straight pipes, different for each beam, merging in a *Y* shaped section just before the *low-beta* defocusing quadrupole. This new layout almost cancels the problems related to beam-beam long range interactions, because the two beams experience only one parasitic crossing inside the defocusing quadrupole where, due to the large horizontal crossing angle, they are very well separated ($\Delta x \sim 20$ $\sigma_x$). It is worth reminding that in the old configuration the colliding beams had 24 parasitic crossing in the IRs and in the main one the separation at the first crossing was in the range $\Delta x \sim 4\div 7$ $\sigma_x$, as can be seen from Fig. 1 and Fig. 6.

## CONCLUSIONS AND ACKNOWLEDGMENTS

Current-carrying wires and octupoles have been used in order to compensate LRBB interactions and crosstalk between beam-beam effects and lattice nonlinearities.

Weak strong simulations proved to be reliable and helpful in finding the proper approach to the compensation of nonlinearities coming from LRBB interaction and from the ring lattice as well.

The wires installed in the DAΦNE IRs proved to be effective in reducing the impact of BBLR interactions and improving the lifetime of the positron beam especially during the KLOE run.

Studying and understanding the impact of the parasitic crossing at DAΦNE had a relevant impact on the definition of the new criteria adopted in the design of the new IRs for the DAΦNE upgrade.

We are indebted to G. Sensolini, R. Zarlenga and F. Iungo for the technical realization of the wires.